\documentclass[final]{svjour2}
\usepackage{graphicx}
\usepackage{rotating}
\usepackage{amssymb}
\usepackage{mathptmx}
\usepackage[square,numbers,compress]{natbib}
\usepackage{tabularx,ragged2e,booktabs,caption}
\usepackage{subfig}
\usepackage{float}
\usepackage{setspace}

\newcolumntype{C}[1]{>{\Centering}m{#1}}

\makeatletter
\journalname{Journal of Low Temperature Physics}

\begin{document}

\newcommand{\hdblarrow}{H\makebox[0.9ex][l]{$\downdownarrows$}-}
\title{Multi-mode TES bolometer optimization for the LSPE-SWIPE instrument}

\author{R. Gualtieri \and E.S. Battistelli \and A. Cruciani \and P. de Bernardis \and M. Biasotti \and D. Corsini \and F. Gatti \and L. Lamagna \and S. Masi}

\institute{Department of Physics, Sapienza, University of Rome,\\P.le Aldo Moro, 5, Rome, 00185, Italy\\ Tel.:+390649914462\\ 
\email{riccardo.gualtieri@uniroma1.it}}

\section*{Acknowledgement}
	The final publication is available at Springer via http://dx.doi.org/10.1007/s10909-015-1436-1

\maketitle
\begin{abstract}
In this paper we explore the possibility of using \emph{transition edge sensor} (TES) detectors in multi-mode configuration in the focal plane of the \emph{Short Wavelength Instrument for the Polarization Explorer} (SWIPE) of the balloon-borne polarimeter \emph{Large Scale Polarization Explorer} (LSPE) for the Cosmic Microwave Background (CMB) polarization.
This study is motivated by the fact that maximizing the sensitivity of TES bolometers, under the augmented background due to the multi-mode design, requires a non trivial choice of detector parameters.
We evaluate the best parameter combination taking into account scanning strategy, noise constraints, saturation power and operating temperature of the cryostat during the flight.

\end{abstract}
\section{Introduction}
	One of the most interesting goals in Cosmology nowadays is the detection of the primordial curl component ("B-mode") of the \emph{Cosmic Microwave Background} (CMB) polarization at large angular scales. This tiny signal might provide the indirect confirmation of the inflationary paradigm since a background of primordial gravitational waves would leave such a characteristic imprint on the CMB. 
	\newline  
	\emph{Transition Edge Sensors} (TES) are the state of the art technology in CMB measurements. Thanks to their high sensitivity they are the choice of cutting-edge experiments both from ground and stratosphere. The readout strategies are mature enough to manage thousands of detectors. 
	\newline
	A step forward for applications which favor	light collection efficiency per detector over spatial resolution, is represented by the use of multi-moded detectors We have modeled and simulated feedhorns and cavities for such purpose as shown in \cite{Lamagna2015}. This allows to significantly increase the sensitivity as for a given number $\mathcal{N}$ of efficiently coupled radiation modes at a given frequency, the photon noise limited $S/N$ ratio improves as $\sqrt{\mathcal{N}}$, at the cost of coarser angular resolution. This is an appealing solution for the LSPE-SWIPE instrument, designed to perform a balloon-borne measurement of CMB polarization at large angular scales \cite{2012arXiv1208.0281T,2012SPIE.8452E..3FD}.
	\newline
	LSPE will observe $25\%$ of the sky in a circumpolar flight during the Arctic night $2016/17$. The SWIPE instrument consists of two focal planes	populated
	with 330 multimoded TES bolometers \cite{multi-mode}, equally split in the 3 bands $140$, $220$, $240$ GHz, and	looking at the sky through a polarizer and cold rotating half wave plate (HWP).
	\newline 	
	This work is motivated by the necessity of optimizing the TES chip under an operating background larger than the one present in standard single-mode detectors operating in the same frequency bands.
	This requires a fine-tuning and trade-off of the characteristics of the TES taking into account their noise and saturation power. 
	In this contribution we present the design of the LSPE-SWIPE detectors in terms of heat capacities $C$ and thermal conductivities $G$ for the chip constituents, and relate them to the bolometer time constant $\tau$.
	\newline
	Finally we discuss the results of the optimization for the specific case of LSPE-SWIPE, for each band of the instrument, constraining the detector parameters as well as the transition temperature of the film $T_C$ and its normal resistance $R_N$. We show that, with these choices, we are able to push the TES technology to its best performance \cite{2014SPIE.9153E..08B}.

\section{Scan Strategy and Time Constant}
	LSPE scans the sky by spinning in azimuth, at a rate $\dot{A}$ , so that the telescope, pointed at elevation $e$, scans the sky at speed $\dot{\theta}\approx \dot{A}\cdot cos(e)$. If the beam has a Gaussian shape with standard deviation $\sigma$ , and the time response of the detector is described by a first-order low-pass with time constant $\tau$ , the degradation of the beam response depends only on the ratio $R$ between the time required to cross one beamwidth and the time constant: 
	\begin{equation}
		R=\frac{\sigma}{\dot{\theta}\tau}=\frac{\sigma}{\tau\dot{A}cos(e)}.
	\end{equation} 
	It has been demonstrated that for $R>2$ the deformation of the beam is negligible for accurate determination of CMB anisotropy and the related cosmological parameters \cite{1998MNRAS.299..653H}.
	\newline	
	In the case of LSPE-SWIPE $\dot{A}\approx3$ rpm, the beam FWHM is $1^\circ.4$ and the minimum elevation is $20^\circ$: the condition $R>2$ results in a requirement for the time constant of the detectors
	 $\tau<15$ ms.
	This can be met for our large throughput bolometers, exploiting the extreme electrothermal feedback conditions achievable with TES sensors.
	Our choice is to consider a time constant:
	\begin{equation}
		\tau=\frac{C}{G(1+\mathcal{L})}\approx10\:ms
	\label{tau}
	\end{equation}
	where $C=7.1\cdot10^{-12}J/K$, that accounts for the chip geometry and the used materials, is the heat capacity of the bolometer,
	$G$ is the thermal dynamic conductance of the link between the sensor and the thermal bath and $\mathcal{L}$ is the loop gain that will be evaluated in what follows. With an assumed transition width $tw=10mK$, a bath temperature $T_{bath}=0.3K$, a critical temperature $T_c=0.5K$ \cite{Biasotti2015} and an $n\sim3.2$, for $Si_3N_4$, we obtain: transition slope $\alpha= T_c/tw = 50$, loop gain $\mathcal{L}=\alpha\Phi/n\simeq13$ where $\Phi\equiv1-(T_{bath}/T_c)^n$ \cite{1998JAP....83.3978I}.
	The constraint on the time constant is also dictated by the choice of the readout scheme. The stability of the resonator, when it is AC biased, in a frequency domain multiplexing environment could be damaged by a small detector time constant \cite{2012RScI...83g3113D}. 
	\newline
	Reverting Eq. (\ref{tau}), this converts into a requirement on the value of $G$ necessary to meet the time constant target such that: $G>50pW/K$.
\section{Noise Constraint}
	In order not have an increase in the Noise Equivalent Power (NEP) larger than $\sim10-15\%$ of photon noise, we aim at a thermal NEP a factor $\sim1.6\div2$ smaller than photon NEP. Photon noises in the three SWIPE bands are: $NEP_{ph}^{140}=6.4\times10^{-17}W/\sqrt{Hz}$, $NEP_{ph}^{220}=5.5\times10^{-17}W/\sqrt{Hz}$ and $NEP_{ph}^{240}=11.5\times10^{-17}W/\sqrt{Hz}$. 
	\newline
	Photon noise has been computed adding in quadrature the contributions from the CMB, Galactic dust at high latitudes, residual atmosphere, cryostat window, optical filters at different temperature stages inside the cryostat. Each contribution has been modeled as a gray-body, with the appropriate emissivity term multiplied by the efficiency of the instrument (filters transmission times bolometer absorption efficiency). For each noise contribution, the Poisson and wave-interference photon noise terms have been added in quadrature. The noise contribution from different radiation modes has been assumed to be uncorrelated. 
	\newline
	Under these assumptions the contribution to the NEP due to photon noise from each source can be computed as:
	\begin{equation}
	\label{NEPph}
	NEP=\sqrt{A\Omega {4k^5 \over c^2h^3} T^5 \int_{x_1}^{x_2} \epsilon_s E {x^4(e^x-1+ \epsilon_s E ) \over (e^x-1)^2} dx }\,\,\,\,\,\,\,\, , \,\,\,\,\,\,  x={h\nu \over kT}
	\end{equation}
	where $E$ is the instrument efficiency while $\epsilon_s$ is the emissivity of the source. This expression could be derived following \cite{1986ApOpt..25..870L}. 
	\newline
	The thermal noise, is defined as: 
	\begin{equation}
		NEP_{th}=\sqrt{4k_BT_c^2GF}
	\label{NEPth}
	\end{equation}	 
	\noindent	
	with $F$ that ranges $0.5\div1$ and takes into account non-equilibrium effects \cite{2005cpd..book...63I}, $k_B$ the Boltzmann constant, $T_c$ the critical temperature and $G$ the conductance. To meet the desired requirements, in each band, we need to have a $NEP_{th}$:
	$NEP_{th}^{140}<4.0\times10^{-17}W/\sqrt{Hz}$, $NEP_{th}^{220}<3.2\times10^{-17}W/\sqrt{Hz}$ and $NEP_{th}^{240}<5.5\times10^{-17}W/\sqrt{Hz}$.
	\noindent
	For a given $T_c$, just as we aim to have a $NEP_{th}<1/2\cdot NEP_{ph}$, by reverting Eq. (\ref{NEPth}) we obtain the desired $Gs$.
	\newline
	\noindent
	In this estimate we are not aware of the thermal fluctuation noise because of the geometry of the chip \cite{2004Tiest}: the thermistor is at the center of a \emph{spider-web} absorber with a radial symmetry, as shown in \cite{Biasotti2015}. 
	
\section{Saturation Power}
\label{SatPow}
	In order not have our TES saturated during observations due the the intrinsic background fluctuations, we set the saturation power requirements $P_{sat} \gtrsim 2.5$ times the power 
	loading expected on the detectors. After accounting for available bandpass information, we calculated the power loading on the detectors as $11$, $5$ and $20$pW respectively for the $140$, $220$, $240$ GHz channels. The power load is evaluated by taking into account the emission of the optics, the bandwidth of the filters ($HPBW$), the atmospheric signal and the throughput.
	The $HPBW$ in each channel is: $HPBW^{140}=33\%$, $HPBW^{220}=5\%$ and $HPBW^{240}=5\%$.
	Those latter narrow bandpasses are chosen to best constraint the slope of the foregrounds spectrum, meeting at the same time the optical load on detector requirement.
	\newline
	With our setup we couple $12$, $30$ and $34$ modes in the $140$, $220$ and $240\:GHz$ bands respectively with an efficiency $\varepsilon$ of $0.3$, $0.2$ and $0.2$.
	\newline
	Modeling the weak thermal link to the heat bath with a temperature dependent conductivity, defined by the law $G=nKT^{n-1}$, with $n \sim 3.2$ for $Si_3N_4$,
	we write $n = 1+\beta$, with $\beta$ the thermal conductivity exponent \cite{2005cpd..book...63I}.
	\newline	  
	The heat flow from the bolometer to the bath can be written as:
	\begin{equation}
		P_{bath}(T)=K(T^n-T_{bath}^n)
	\end{equation}
	where $n$ and $K$ are relative to the particular bolometer and material configuration. Since the saturation power is the power dissipated to the bath at the superconducting temperature we can write:
	\begin{equation}
		P_{sat}(T_c)=K(T_c^n-T_{bath}^n)
	\label{Psat}
	\end{equation}
	where everything depends on the thermal conductance $G$ between the bolometer and the bath:
	\begin{equation}
		G=\frac{dP_{bath}}{dT}\Big|_{T_c}=nKT_c^{n-1}
	\end{equation}
		
\section{Thermal conductance vs critical temperature}
	The fridge base temperature is expected to be around $300mK$. The required $T_c$, taking into account unavoidable temperature gradients and non-uniformities will be at the level of	$450mK-500mK$.
	From what described in the previous sections, we can set three different constraints on the bolometer conductivity $G$ as a function of the TES critical temperature $T_c$, with a loop gain $\mathcal{L}=13$ and a $C=7.1\cdot10^{-12}J/K$ that accounts for the chip geometry and the materials:
	
	\noindent\begin{tabularx}{\textwidth}{XXX}
	\bigskip
	From Eq. (\ref{NEPth}):
		\begin{equation}
			G^{NEP}\leq\frac{NEP^2}{4Fk_BT_c^2}
		\label{GNEP}
		\end{equation} &
	\bigskip
	From Eq. (\ref{Psat}):
		\begin{equation}
			G^{P_{sat}}\geq\frac{nP_{sat}T_c^{n-1}}{T_c^n-T_{bath}^n}
		\label{GPsat}
		\end{equation} &
	\bigskip
	From Eq. (\ref{tau}):
		\begin{equation}
			G^{\tau}\geq\frac{C}{\tau(1+\mathcal{L})}
		\label{Gtau}
		\end{equation}
	\end{tabularx}
	\newline
	Each inequality defines the space parameters in which noise performance of the system are preserved (\ref{GNEP}), the experiment base temperature and power on the detectors is taken into account (\ref{GPsat}) as well as the detector's time constant, constrained by the scanning strategy (\ref{Gtau}). 
	
	\begin{figure} 
	\centering
	    \subfloat[\label{subfig-1:Gs_140}]{%
	      \includegraphics[width=0.8\textwidth]{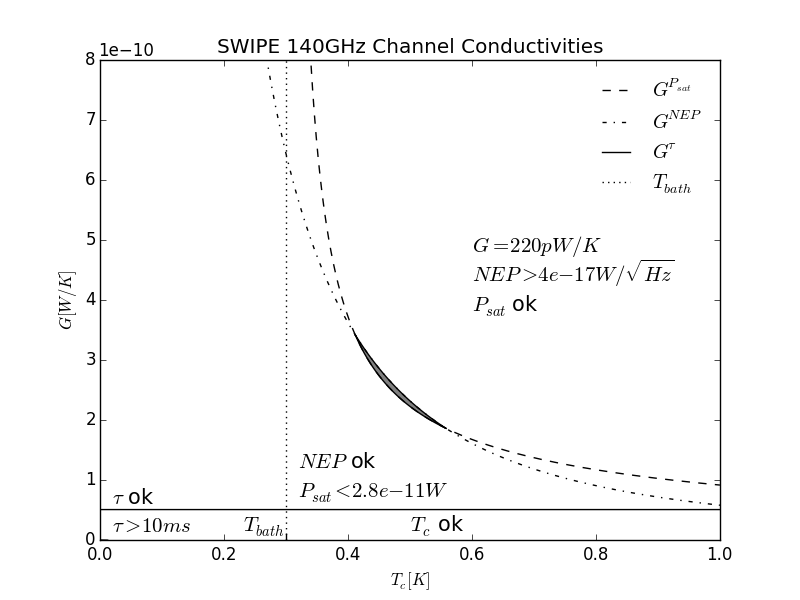}
	    } \qquad\qquad
	    \subfloat[\label{subfig-2:Gs_220}]{%
	      \includegraphics[width=0.8\textwidth]{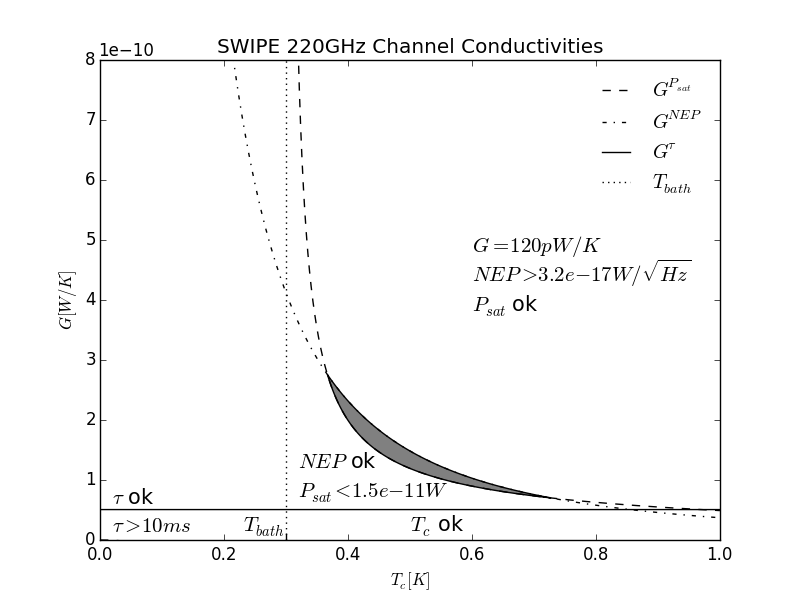}
	    } \qquad\qquad
	    \subfloat[\label{subfig-3:Gs_240}]{%
	    	\includegraphics[width=0.8\textwidth]{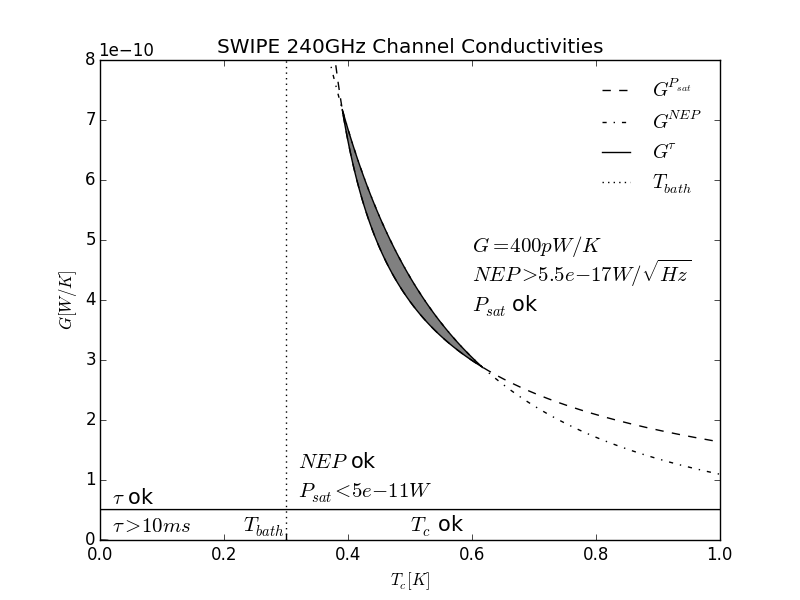}
	    }
	    \caption{The plots in figures represent the space of the parameters of the bolometers in therms of their conductances and critical temperature. The calculation is made for the three channels of the SWIPE instrument: Fig.~\ref{subfig-1:Gs_140}, Fig.~\ref{subfig-2:Gs_220} and Fig.~\ref{subfig-3:Gs_240} respectively. The shaded regions indicate the allowed parameter space.}
	    \label{fig:Plot_Res}
	\end{figure}

\section{Results and Conclusions}
	The allowed parameters space for $G$ and $T_c$ is depicted as shaded regions in plots of Fig.~\ref{fig:Plot_Res} and the optimal values for the parameters are reported in Table \ref{tab:Results}.
	For $P_{sat}$, $\tau$ the allowed regions are on the upper side of the curves $G^{P_{sat}}$ and $G^\tau$ respectively. For the $NEP$ the allowed region is the lower side of the $G^{NEP}$ curve.
	From fig.\ref{fig:Plot_Res} is evident that the acceptable size of the parameter space is small.
	This means that the TES bolometers construction parameters need to be fine-tuned to the incident power load with even more care as for single mode TES.
	\newline
	Just as we adopt multimode optics to collect the radiation, the incident power is higher than single mode ones. 
	In the specific case of the multi-moded detectors of LSPE, we have demonstrated with this study that it is indeed possible to find	a solution optimizing TES detectors performance, but this requires exquisite
	control of fabrication parameters and a very careful model of the power load on the detectors in all the phases of detector use. 
	
	\begin{minipage}{0.96\linewidth}
		\bigskip
		\captionof{table}{TES bolometers requirements as calculated with our model. Taking into account of all hypothesis and constraints we have chosen the best values for the parameters.} \label{tab:Results}	
		\begin{tabularx}{0.95\linewidth} {C{0.25\linewidth}*4X} \toprule[1.5pt]
		Frequency [GHz] & $140$ & $220$ & $240$\\\midrule
		$\tau[ms]$ & $10$ & $10$ & $10$\\ \midrule
		$T_c[mK]$ & $500$ & $500$ & $500$\\ \midrule
		$tw[mK]$ & $10$ & $10$ & $10$\\ \midrule
		$HPBW$ & $33\%$ & $5\%$ & $5\%$ \\\midrule 
		$\mathcal{N}_{modes}$ & $12$ & $30$ & $34$ \\\midrule
		$\varepsilon$ & $0.3$ & $0.2$ & $0.2$ \\\midrule
		$NEP_{ph}[W/\sqrt{Hz}]$ & $6.4\times10^{-17}$ & $5.5\times10^{-17}$ & $11.5\times10^{-17}$\\\midrule
		$NEP_{th}[W/\sqrt{Hz}]$ & $4.0\times10^{-17}$ & $3.2\times10^{-17}$ & $5.5\times10^{-17}$\\ \midrule
		$P_{opt}[pW]$ & $11$ & $5$ & $20$\\ \midrule
		$P_{sat}[pW]$ & $28$ & $15$ & $50$\\ \midrule
		$G[pW/K]$ & $220$ & $120$ & $400$\\ 
		\bottomrule[1.25pt]
		\end {tabularx} \par 
		\bigskip
	\end{minipage}

\pagebreak

\end{document}